\documentstyle[psfig]{mn}
\title{The anticenter old open cluster  NGC~1883: radial velocity and metallicity}
\author[Villanova  at al.]
{Sandro Villanova$^{1}$,
 Gustavo Baume$^{2}$,
 Giovanni Carraro$^{1}$
 \thanks{email: sandro.villanova@unipd.it (I),
                gbaume@fcaglp.fcaglp.unlp.edu.ar (Ar),
 giovanni.carraro@unipd.it (I)}\\
      $^1$Dipartimento di Astronomia, Universti\`a di Padova, vic. Osservatorio 3,
Padova, Italy\\
      $^2$Facultad de Ciencias Astron\'omicas y Geof\'{\i}sicas de la UNLP,
IALP-CONICET, Paseo del Bosque s/n, La Plata, Argentina \\
}

\date{\it Submitted: -------}
\pubyear{2007}
\begin{document}
\maketitle
\title{}

\begin{abstract}
Having already reported on the first photometric study of the
intermediate-age open cluster NGC~1883 (Carraro et al. 2003),
we present in this paper
the first spectroscopic multi-epoch investigation of a sample of evolved
stars in the same cluster.
The aim is to derive the cluster membership, velocity and metallicity, and
discuss recent claims in the literature (Tadross 2005)
that NGC~1883 is as metal poor as globular clusters in the
Halo.
Besides, being one of the few outer Galactic disk intermediate-age open clusters
known so far,
it is an ideal target to improve our knowledge of  the Galactic disk
radial abundance gradient, that is a basic ingredient for any chemical
evolution model of the Milky Way.
The new data we obtained allow us to put NGC 1883's 
basic parameters more reliable. We find that the cluster has a mean
metallicity of [Fe/H] = -0.20$\pm$0.22,
from which we infer an age (650$^{+70}_{-70}$ Myr) close to the Hyades one
and a Galactocentric distance of 12.3$^{+0.4}_{-0.2}$ kpc.
The metal abundance, kinematics, and position make NGC 1883 a genuine
outer disk intermediate-age open cluster. We confirm that in the outer
Galactic disk the abundance gradient is shallower than in the solar
vicinity.
\end{abstract}

\begin{keywords}
Open clusters and associations: general-- Open clusters and
associations: individual: NGC1883-- Milky Way: structure and evolution
\end{keywords}

%
\section{Introduction}
%
NGC 1883 is a northern open cluster, located in the second Galactic
quadrant, toward the anticenter
direction ($\alpha=05^h25^m.9,\delta=+46^029^{\prime},
\ l=163^0.08,b=+06^0.16,J2000.0$).
This cluster was studied by Carraro et al. (2003, hereafter C03),
who present the first BVI CCD photometric observations
suggesting an age of about 1 Gyr and a distance of 4.8 kpc from the
Sun.\\
This would imply a Galactocentric distance larger than 13 kpc, and
would make NGC 1883 a key object four our understanding of the
properties of the outer Galactic disk.
However, both age and distance
are uncertain because of the unknown metallicity of the cluster. \\
At the position of NGC 1883 the mean abundance
(Friel 1995) should be significantly lower than solar.
Unfortunately,  the exact amount of this under-abundance is
poorly constrained, also in light of
the results by Twarog et al. (1997),
Carraro et al. (2004),  and Yong et al. (2005), that  suggest
that the  disk abundance gradient for Galactocentric distances
larger than $\sim$10 kpc gets much shallower than in the solar
vicinity.\\
Using solar metallicity (Z=0.019) C03 estimated E(B-V)=0.23,
(m-M)$_V$=14.00 for NGC 1883, and E(B-V)=0.35,
(m-M)$_V$=14.50 adopting a lower metal content
(Z=0.008), which by the way seems to provide a better
fit to the star distribution in the Colour Magnitude Diagram (CMD).\\
A more recent study was performed by Tadross (2005, hereafter T05)
using C03 BVI data and JHK 2MASS data.\\
This author obtained a photometric metallicity of about [Fe/H]$\sim$-1.1
(Z$\sim$0.0015), much lower than that estimated by C03, and closer
to globular clusters.
The disagreement between the two abundance
estimations has to  be solved in order to obtain
reliable parameters for this cluster.
In fact,
NCG 1883 is one of the few intermediate-age open clusters located in the
outer Galactic disk.
For this reason, it plays an important role in defining the shape of the radial
abundance gradient in the external regions of the Galactic
disk, which is one of the fundamental costrains of models of chemical
evolution (Cescutti et al. 2007).\\
For this purpose, this paper presents the first spectroscopic study
of the cluster aiming at finding the member stars from radial velocity
measurements, and to estimate a reliable value for the metallicity.
Section 2 of this paper illustrates the observation and reduction
strategies. In Section 3 and 4 we discuss the membership and the
abundance determination.
Section 5 is devoted to a comparison between this paper
and previous ones.
Section 6 deals with a new determination of cluster reddening,
distance and age.
Section 7 compares the observed CMD with a synthetic one.
Section 8 and 9 are devoted to the determination of structural
parameters for the cluster (radial profile, luminosity function, and
mass function).
Finally Section 10 discusses NGC1833 in the context of the Galactic abundance
gradient.

\begin{table*}
\caption{Observed stars}
\label{tab1}
\centering
\begin{tabular}{lccccccccc}
\hline\hline
ID(U-1350-057...) & R.A.(J2000.0) & Decl.(J2000.0) & V & B-V & V-I & V-K & \small{${\rm RV}_{\rm{H}}$(A)}  & \small{${\rm RV}_{\rm{H}}$(B)} & \small{${\rm RV}_{\rm{H}}$(C)}\\
\hline
79024  & 05:26:02.01 & +46:29:26.9 & 12.30 & 1.85 & 1.89 & 4.32 &  34.4 &  33.0 & 34.4 \\
75374  & 05:25:51.69 & +46:29:35.4 & 13.25 & 1.78 & 1.79 & 4.00 & -29.7 & -32.5 & -32.1\\
76058  & 05:25:53.66 & +46:26:20.5 & 13.30 & 1.61 & 1.70 & 3.70 & -29.8 & -30.3 & -30.7\\
81797  & 05:26:09.92 & +46:25:55.4 & 11.85 & 1.87 & 1.94 & 4.38 &   2.7 &  3.0  &   2.7\\
85248  & 05:26:19.92 & +46:31:41.4 & 12.61 & 2.18 & 2.60 & 5.43 & -17.9 & -18.5 & -18.4\\
\hline
\end{tabular}
\end{table*}

%
\section{Observations and data reduction}
%

Medium-high resolution ($R \approx 20000$) spectra
of 5 candidate evolved stars in the field
of NCG1883 have been obtained using the REOSC Echelle Spectrograph
on board of the 1.82 m telescope of Asiago Astronomical Observatory.
The Echelle spectrograph works with a Thomson 1024 $\times $ 1024 CCD
and the wavelength coverage for our observations is approximately 4600-6400
\AA. Details on this instrument are given in the Asiago Observatory
Home page (www.oapd.inaf.it/asiago/).\\
The exposures were 3$\times$15 min for all stars. To improve the signal
to noise ratio the three exposures were added reaching S/N
values up to 20.
Each star was observed during three epochs, December 2004 (A),
January 2005 (B), January 2006 (C), allowing us
to establish the binarity of each target.

\begin{figure}
\centerline{\psfig{file=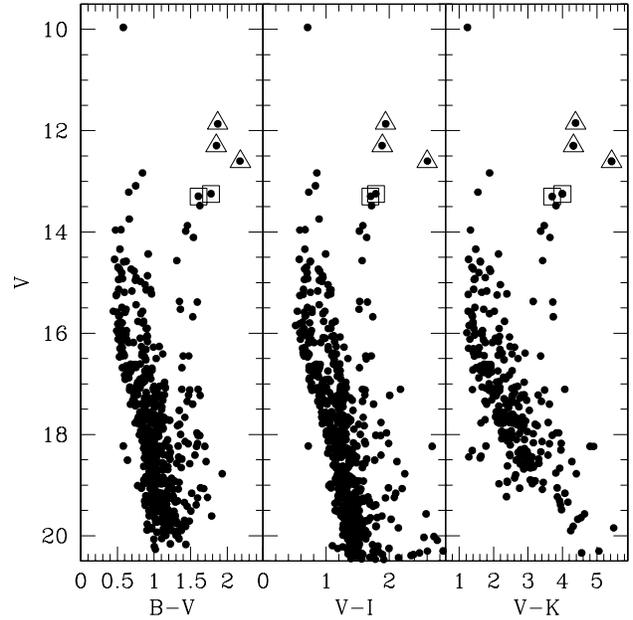,width=9cm}}
\caption{CMD for the stars in the field of NGC 1883.
         {\it left:} V vs B-V diagram from C03. {\it center:}
         V vs V-I diagram from C03. {\it right:}
         V vs V-K diagram (K magnitude from 2MASS). Open squares are
         member stars (\#75374,\#76058), while open triangles
         are background objects. See text for more details.}
\label{fig1}
\end{figure}

Data were reduced with the IRAF package\footnote{IRAF is distributed
by the National Optical Astronomy Observatory, which is operated by
the Association of Universities for Research in Astronomy, Inc., under
cooperative agreement with the National Science Foundation.},
including bias subtraction, flat-field correction, frame combination, extraction of
spectral orders, wavelength calibration, sky subtraction, and spectral
normalization.
By comparing known sky lines positions along the spectra we obtained an
error in wavelength calibration of less than 0.01 \AA\ ($<$0.5 km/s).\\
Details of the observed stars are listed in Table~\ref{tab1}, while
the positions of these targets on CMDs are plotted in Fig.~\ref{fig1}.
The first column of Tab.~\ref{tab1} gives the ID number of the star
according to USNO-2.0 catalogue, the second and third
the coordinates, the fourth, fifth, sixth, and seventh the V magnitude
and B-V,V-I,V-K colours, the eight, ninth, and tenth columns the
measured radial velocity at different epochs expressed in km/s.\\
The photometry for our stars was taken from C03 (B,V,I magnitudes),
and from 2MASS (K magnitude).

Radial velocity was measured using the IRAF package FXCOR, which
cross-correlates the observed spectrum with a template.
As a template we used a synthetic spectrum having mean atmospheric
parameters close to the cluster RGB stars (T$_{\rm eff}$=4500 K,
log(g)=3.0, [Fe/H]=0.0, v$_t$=1.5 km/s).
Finally the heliocentric correction was applied.
The typical error for our measurements is less than 1 km/s.

%
\section{Membership and Cluster Mean Radial Velocity}
%

We make use of these radial velocities to
establish star membership to NGC 1883.
Firstly, we notice that each star has radial velocities constant
among the three epochs within the errors.
We are led to the conclusion that none of the targets are
short-period binaries.
The mean radial velocity and membership of all the targets
is reported in Tab.~\ref{tab2} where
member stars are indicated by M, while non-member by NM.
Only \#75374 and \#76058 stars have compatible radial velocity
and therefore we consider them as genuine members (open
squares in Fig.~\ref{fig1}) of the cluster.
The average of the velocities of \#75374 and \#76058 gives us the
mean radial velocity of the cluster:
\ \\
\begin{center}
$<RV_{helio}>=-30.8\pm0.6\ {\rm km/s}$
\end{center}
\ \\
The members stars will be used in the following Section
to obtain the metal content of NGC 1883.

%
\section{Abundance measurements}
%

The [Fe/H] content was obtained from summed spectra
of member stars (\#75374 and \#76058) in a interactive way.
The most important parameter for abundance
determination is T$_{\rm eff}$, that for our stars can be obtained
from dereddened B-V, V-I, and V-K colours. But the reddenings E(B-V),
E(V-I), and E(V-K) for the cluster
can be obtained only by isochrones fitting, with isochrones
having the correct metallicity.
For this reason we firstly assumed a solar metallicity
for our stars, obtaining a first guess for reddening
by isochrones fitting. Then, using the T$_{\rm eff}$ value
calculated for this reddening, we measured the metallicity
for our stars (see below). Using this metallicity,
we obtained a new reddening and a new abundance. This procedure was iterated
until convergence was achieved.

T$_{\rm eff}$ was derived from the colour-[Fe/H]-temperature
relation from Alonso et al. (1999) using dereddened colours.
The gravity was derived from the value given by the isochrones
at the given position of the stars in the CMD, while
the microturbulent velocity come from the relation (Houdashelt et
al. 2000):
\ \\
\begin{center}
$v_{\rm t}=2.22-0.322 \times {\rm log(g)}$
\end{center}
\ \\
The adopted atmospheric parameters for our stars are reported in
Table~\ref{tab2}, and the typical random errors are 100 K for T$_{\rm eff}$,
0.2 for log(g), and 0.2 km/s for v$_{\rm t}$.\\
We derived metallicity from equivalent widths of selected spectral
lines. The equivalent width was derived by gaussian fitting
of these spectral features. Repeated measurements show a typical error of
about 10 m\AA. Because of the low S/N and the resolution of our
spectra, we could measure only the strongest Fe lines ($\sim$10 for
each star) in the region $\lambda>5500$ to avoid crowding.
For this reason we give only Fe content of our stars (see
Tab.~\ref{tab2}).
The LTE abundance program MOOG (freely distributed by C. Sneden, 
University of Texas, Austin; see http://verdi.as.utexas.edu/) was 
used to determine the metal abundances. Model atmospheres were 
interpolated from the grid of Kurucz (1992) models by using 
the values of Teff and log(g) of Tab.~\ref{tab2}. 
The error on measurements is high (0.2-0.4 dex), but a weighted
mean of the obtained metallicities gives a good estimation
of the iron content:
\ \\
\begin{center}
$[{\rm Fe/H}]=-0.20\pm0.22$
\end{center}
\ \\
This value is important for the discussion reported
in the following Sections and for the comparison
with previous works.
The errors for iron content we report here and in Tab.~\ref{tab2} represent only internal random errors,
mainly due to the uncertainty in the equivalent width measurements.
Systematic errors are more difficult to estimate, and are mainly due
to uncertainties in colour-T$_{\rm eff}$ relations
and in reddening. The typical systematic error on T$_{\rm eff}$ scale
for this temperature range is 100 K giving
a systematic uncertainty on metallicity of 0.02 dex.

\begin{table*}
\caption{Mean radial velocity, membership, and parameters for the observed stars}
\label{tab2}
\centering
\begin{tabular}{lcccccc}
\hline\hline
ID & $<$RV$_{\rm{H}}$$>$ (km/s) & Membership & T$_{\rm eff}$ (K) & log(g) & v$_{\rm t}$
(km/s) & [Fe/H]\\
\hline
79024  &  33.9 & NM &   -  &   -  &  -  &        -      \\
75374  & -31.4 & M  & 4172 & 2.75 & 1.3 & -0.25$\pm$0.26\\
76058  & -30.3 & M  & 4462 & 2.75 & 1.3 & -0.07$\pm$0.46\\
81797  &   2.8 & NM &   -  &   -  &  -  &        -      \\
85248  & -18.3 & NM &   -  &   -  &  -  &        -      \\
\hline
\end{tabular}
\end{table*}

%
\section{Comparison with previous work}
%

Firstly, we compare our results with C03.
In that paper the authors estimate a
range in metallicity between Z$\sim$0.019 ([Fe/H]=0.0)
and Z$\sim$0.008 ([Fe/H]=-0.41) by isochrones fitting
of the TO.
Our measurements lie exactly in the middle confirming
those results.
On the other hand, our value largely disagrees
with T05. This author gives [Fe/H]$\sim$-1.1,
based on the slope of the RGB stars (Vallenari et al. 2000 and
references therein).
The method can give a reasonable estimate
of the metallicity of a cluster (Tiede et al. 1997), but [Fe/H]$\sim$-1.1
is  definitely ruled out by  our measurements.
T05 results can be explained by a poor treatment of
stellar contamination.
As a matter of fact, background stars always contaminate
the field of a cluster, and an identification of the
members is needed before applying RGB-slope method.
This is the case of NGC 1883,
where a fraction of stars that apparently populate
the RGB are background objects, as  Fig.~\ref{fig1}
unambiguously shows.
This contamination changes the real slope of the RGB and
the [Fe/H] estimate by T05 is clearly affected by this problem.
As shown in Fig.~\ref{fig1} the RGB slope in T05 is mainly
based on \#79024,\#81797, and \#85248 stars (open triangles) that, according to
our radial velocities, are not members.
This leads us to the conclusion that T05 metallicity
is untenable.

%
\section{Cluster fundamental parameters}
\begin{figure}
\centerline{\psfig{file=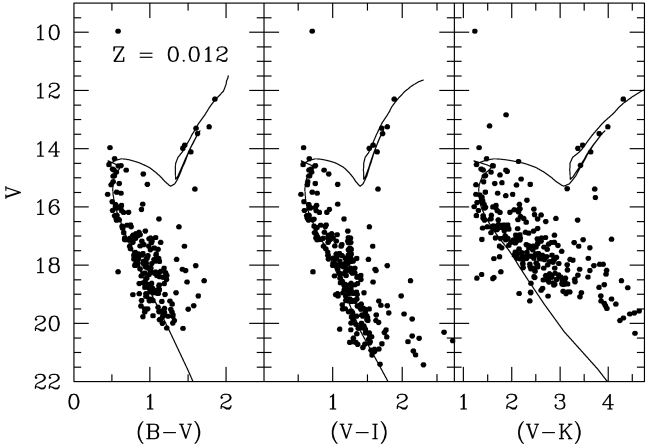,width=9cm}}
\caption{Isochrones fitting of the photometric data using
         the metallicity obtained in Section 4.
         {\it left:} V vs B-V diagram from C03. {\it center:}
         V vs V-I diagram from C03. {\it right:}
         V vs V-K diagram (K magnitude from 2MASS).
         See text for more details.}
\label{fig2}
\end{figure}

%
Having an estimate of the metal content ([Fe/H]=-0.20),
we now provide a more reliable estimate
of the cluster parameters.
We make use of the comparison between the distribution of
the stars in the CMDs presented in C03 and a set of theoretical
isochrones from the Padova group (Girardi et al. 2000).
In Fig.~\ref{fig2} we superimposed an isochrone with [Fe/H]=-0.20
(Z=0.012) and an age of 650 Myr.
The TO fit is very good both in V vs B-V, V vs V-I, and V vs V-K diagrams,
and the shape of the MS is well reproduced.
Also the location of \#75374 and \#76058 stars in the RGB is well matched.
By this fitting we obtained (m-M)$_{V}$=14.30 and E(B-V)=0.43,
E(V-I)=0.52, E(V-K)=1.15.\\
The reddening value we obtained is compatible with Schlegel et
al. (1998) maps which, in the direction of NGC 1883, predict
a E(B-V) value of 0.55, all the way to infinity.

NGC 1883 turns out to be located 3.9 kpc from the Sun
towards the anticenter direction. This yields from the
heliocentric rectangular Galactic coordinates X=-3.71 kpc, Y=1.13 kpc, Z=0.41 kpc, and
a distance from the Galactic center of 12.3 kpc.\\
The age we found in this study is somewhat smaller (about 350 Myr) than in C03,
but NGC 1883 can still be considered one of the few
intermediate-age open cluster located in the outer Galactic disk.

The distance estimate is also confirmed by the V magnitude
of the red-clump stars (Girardi \& Salaris 2001).
They are identified with the group of three stars
having $<$V$>$=14.0 and $<$(B-V)$>$=1.45 (see Fig.~\ref{fig2}).
A V magnitude of 14.0 for red-clump stars gives
a distance of 3.6$\pm$0.4 kpc from the Sun, that
agrees well with the value we obtained with isochrone fitting.

We interpret the two member stars at V 13.3 as bright RGB stars.
In fact, the interpretation of these two stars as clump stars would
imply a magnitude difference between clump and TO  $\Delta V = 2.1$,
and, in turn, an age larger than 3 Gyr (Carraro \& Chiosi 19944).
An isochrone for this combination of age and metallicity would have
too a blue clump, and produce a serious mismatch of the evolved stars
distribution. Besides, it would imply a reddening close to zero
(E(B-V) $\sim$ 0.06), too small for a cluster which in this case would lie 
at 1.7 kpc from the Sun.

\section{Synthetic CMDs}
\begin{figure}
\centerline{\psfig{file=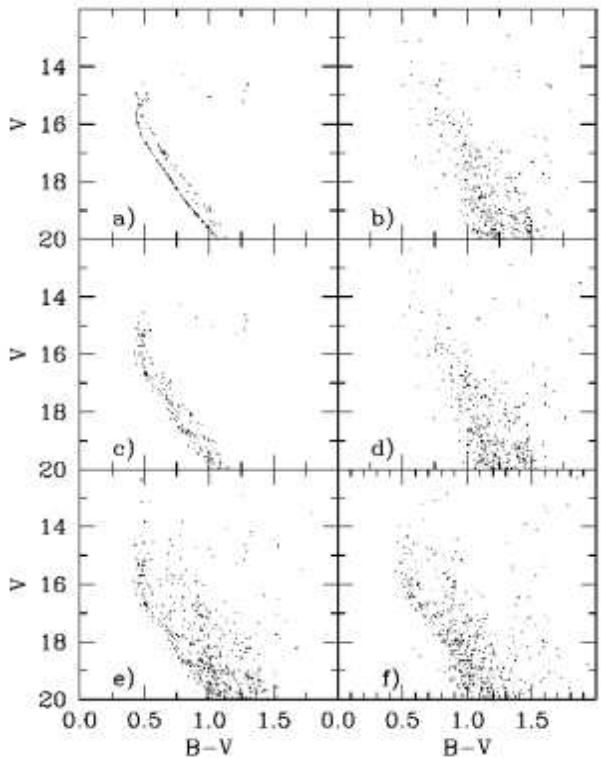,width=\columnwidth}}
\caption{Synthetic CMD diagram for NGC 1883. Panel a) and b):
         simulated CMDs for the cluster and the field.
         Panel c) and d): simulated CMDs for the cluster and the field with
         photometric errors
         Panel e): final simulated CMD (cluster+field).
         Panel d): the observed CMD.}
\label{fig3}
\end{figure}

To better establish the basic parameters of the cluster we calculated
synthetic $V$ vs $B-V$ CMDs.
We started from the Girardi et al.\ (2000)
theoretical isochrone of metallicity $Z=0.012$ used for our previous
fitting. The isochrone was shifted
in apparent magnitude and colour of the amount found in the last
Section.
The sequence of steps required to simulate a CMD aimed to
reproduce the NGC~1883 data are:
    \begin{itemize}
    \item
The 650 Myr old isochrone of $Z=0.012$ is used to simulate a
cluster with 3 red clump stars (see previous Section).
Assuming a Kroupa (2001) IMF,
in order to reach this number we need an initial cluster mass of
about $8.0\times10^2$~$M_\odot$, which is assumed herein-after.
According to the same IMF, we would expect 9-11 stars located
beyond the subgiant branch region all the way to the
asymptotic giant branch (AGB) region. The exact number and
distribution of these stars depend on the simulation {\it seed}. Anyhow,
this range of number matches the observations.
In fact (see Fig.~2), despite the fact that
always some contamination is expected, we count 7$\pm$3 evolved stars,
taking into account Poisson uncertainties.

We have simulated detached binaries, assuming that 30 percent
of the observed objects are binaries with a mass ratio
between 0.7 and 1.0. This prescription is in agreement with
several estimates for galactic open clusters (Phelps \& Janes 1994)
and with the observational data for NGC~1818 and NGC~1866 in
the LMC (Elson et al.\ 1998; Barmina et al. 2002).
The result of such simulation is shown in Fig.~\ref{fig3}(a).
In this panel we can see that most of the stars -- the
single ones -- distribute along the very thin sequence defined
by the theoretical isochrone. Binaries appear as both (i) a sequence of
objects roughly parallel to the main sequence of single stars, and
(ii) some more scattered objects in the evolved part of the CMD.
    \item
In order to estimate the location of foreground and background
stars, we use the TRILEGAL\footnote{http://trilegal.ster.kuleuven.be/cgi-bin/trilegal} Galaxy model code
(Girardi et al. 2005). It includes
the several Galactic components -- thin and thick disk, halo,
and an extinction layer -- adopting geometric parameters as
calibrated by Groenewegen et al.\ (2001) (for other applications
see Baume et al. 2007, Carraro et al. 2006).
The most relevant component in this case is the thin disk, which is modelled by
exponential density distributions in both vertical and radial
directions. The radial scale heigth is kept fixed (2.8 kpc),
whereas the vertical scale heigth $h_z$ increases with the stellar
age $t$ as

\begin{center}
$h_z = z_0 (1+t/t_0)^\alpha$
\end{center}

with $z_0=95$ pc, $t_0=4.4$ Gyr, $\alpha=1.66$.
The simulated field has the same
area ($8.1 \times 8.1\, {\rm arcmin}^{2}$) and galactic coordinates
($\ell=163^\circ.08$, $b=+6^\circ.16$) as the observed one for
NGC~1883. The results are shown in Fig.~\ref{fig3}(b).
It is noteworthy that, in this direction, most of the Galactic
field stars appear in a sort of diagonal sequence in the CMD.
    \item

Within TRILEGAL, there are various options to take reddening into
account (Girardi et al. 2005). In the particular case of the
simulation we are presenting here, we adopt an exponential dependence
of reddening with distance, calibrated at infinity with Schlegel et
al. (1998) maps, which provide the reddening along a line of sight all the way
to infinity.
\item

We then simulate the photometric errors as a function of $V$
magnitude, with typical values derived from our observations.
The results are shown separately for cluster
and field stars in panels (c) and (d) of Fig.~\ref{fig3}.
    \item
The sum of field and cluster simulations is shown in
Fig.~\ref{fig3}(e). This can be compared directly to the
observed data shown in Fig.~\ref{fig3}(f).
    \end{itemize}
The comparison of these two latter panels indicates that the selected
cluster parameters -- age, metallicity, mass, distance, reddening, and
binary fraction -- really lead to an excellent description of the
observed CMD, when coupled with the simulated Galactic field.
The most noteworthy aspects in this comparison are the location and
shape of the turn-off and subgiant branch, that
are the features most sensitive to the cluster age.

Of course, there are minor discrepancies between the observed
and simulated data, namely:
(i) The simulated cluster is better delineated in the CMD than
the data. This may be ascribed to a possible underestimate of
the photometric errors in our simulations, and to the possible
presence of differential reddening across the cluster
(ii) There is a deficit of simulated field stars, that can be noticed more clearly
for $V<16$ and $(B-V)<1$. This is caused by the simplified
way in which the thin disk is included in the Galactic model: it is
represented by means of simple exponentially-decreasing stellar
densities in both radial and vertical directions, and does not
include features such as spiral arms, intervening clusters, etc.,
that are necessary to correctly describe fields at low
galactic latitudes. Anyway, the foreground/background simulation we
present is only meant to give us an idea of the expected location of
field stars in the CMD.

Although these shortcomings in our simulations might
probably be eliminated with the use of slightly different
prescriptions, they do not affect our main results, that regard the
choice of cluster parameters
Then, we conclude that (m-M)$_V=14.3$, $E_{B-V}=0.43$,
650~Myr, $Z=0.012$ ($[{\rm Fe/H}]=-0.20$)
well represent the cluster parameters. All these values are
uncertain to some extent:
    \begin{itemize}
    \item
 From isochrones fitting we can estimate a maximum error of
10 percent (70~Myr) in the age. One should keep in mind, however,
that the absolute age value we derived depends on the
choice of evolutionary models, and specially on the prescription
for the extent of convective cores. For the stellar masses involved
($M_{\rm TO} \sim 1.8$~$M_\odot$ for NGC~1883),
our models (Girardi et al. 2000) include a moderate amount of core
overshooting.
    \item
Our best fit model corresponds to E(B-V)=0.43 and
(m-M)$_V$=14.30, but typical errors for both measurements
can reach 10-15\%.
    \item
The initial mass estimate depends heavily on the choice of
IMF, that determines the mass fraction locked into low-mass
(unobserved) objects. The value of $8.0\times10^2$~$M_\odot$ was obtained
with a Kroupa (2001) IMF, corrected in the lowest mass interval
according to Chabrier (2001; details are given in Groenewegen et al.
2001), and should be considered just as a first guess.
At present ages, supernovae explosions, stellar mass loss,
and tidal effects (Lamers et al. 2005) would have reduced
this mass by about 40 percent.
    \end{itemize}

\section{Surface density profile}
\begin{figure}
\centerline{\psfig{file=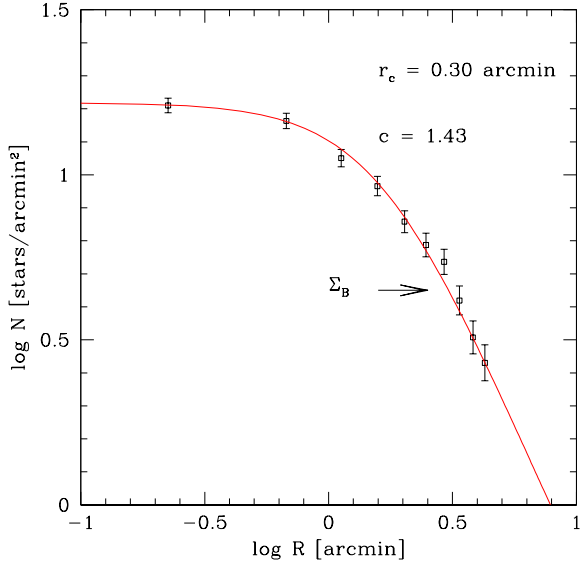,width=8cm}}
\caption{Surface density profile for the cluster.
         The solid line is a King model for the indicated parameters.
         $\Sigma_B$ shows the level of the field background.}
\label{fig4}
\end{figure}

According to C03 and Dias et al. (2002), NGC 1883 has a radius
of 2.5-3.0 arcmin. This is the distance from the cluster
center where star counts remain flat, reaching the level of the Galactic
field. However, to provide  more physical quantities, we perform here
a comparison with King (1962)  models (Fig.~\ref{fig4}).\\
To this aim we first adopt a new cluster center based on the stellar concentration.
C03 chose star $\#$12 as center (see Fig.~\ref{fig5}), but a more detailed inspection
of the cluster appearance suggests that a more appropriate
option for the center is  X = 495, Y = 605 pixels (see
Fig.~\ref{fig5}), which in turns is ($\alpha_{2000} = 05^h25^m58^s$;
$\delta_{2000} = +46^029^{\prime}03^{\prime\prime}$).\\
By adopting the new cluster center, we counted the number of
stars as a function of the distance from the center, and correct
these counts by background contamination. The choice
of the background is illustrated in Fig.~\ref{fig5}, and implies that
one can expect to encounter $\sim$ 4.5 stars/arcmin$^2$ in the
direction of NGC~1883 down to V$\sim$20. This value is in reasonable
agreement with the predictions of the TRILEGAL model (Section 7),
according to which we expect to find 5.2 stars/arcmin$^2$ towards
at (l=163.1,b=+6.2).\\
\noindent
In Fig.~\ref{fig4} we show star counts
together with their Poisson uncertainties.
The solid line is the best-fit King profile, which yields
a core radius $r_c$ = 0.30 arcmin and a concentration
$c = log(\frac{r_t}{r_c})$ = 1.43, where $r_t$ is the cluster
tidal radius. $r_t$ results 8.1 arcmin, much larger than the region
covered by our photometry.
These values can be compared with the recent study by Piskunov et al. (2007),
where the authors study the $r_c$ and $r_t$ distribution of a sample
of open clusters in the Sun vicinity.
According to this study, the open clusters mean $c$ value is
$\sim$0.6. Therefore, NGC~1883 looks like a poorly concentrated cluster, on the verge
of dissolving into the Galactic disk general field.

\section{Cluster luminosity and mass function}
With the aim of deriving an estimate of NGC~1883 integrated
luminosity and mass, we construct the luminosity
function (LF) of the cluster.\\

\begin{figure}
\centerline{\psfig{file=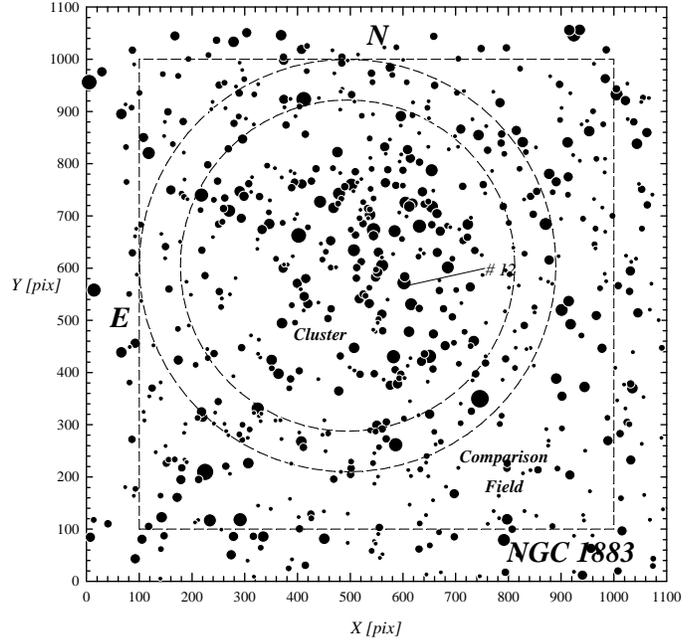,width=9cm}}
\caption{The field used for the Cluster Luminosity Function
         determination. The {\it Cluster Region} and the {\it Comparison Field}
         are indicated.}
\label{fig5}
\end{figure}

\noindent
As a first step, we considered the
$Cluster~Region$ and the $Comparson~Field$: the first one
is a circle centered in the adopted cluster center and $2\farcm5$
in radius; and the later is a region limited for a circle of
$3\farcm11$ in radius and a $7\farcm1 \times 7\farcm1$ square
(see Fig.~\ref{fig5}).
Both regions were chosen to cover the same area.

We estimated then the completeness of our data as in Baume et al. (2004)
and assume that it is the same across the whole region.
This seems to be a good approximation, provided the cluster
is not very crowded.
The completeness as a function of magnitude $V$ is given
in Table~\ref{tab3}

\begin{table}
 \fontsize{8} {10pt}\selectfont
 \tabcolsep 0.1truecm
 \caption{Completeness results  in NGC 1883.}
 \begin{center}
 \begin{tabular}{cc}
 \hline
 \multicolumn{1}{c}{$V$}         &
 \multicolumn{1}{c}{\%}       \\
 \hline
10.5 & 100.0\\
11.5 & 100.0\\
12.5 &  99.9\\
13.5 &  99.5\\
14.5 &  97.3\\
15.5 &  96.1\\
16.5 &  95.8\\
17.5 &  90.1\\
18.5 &  73.4\\
19.5 &  62.0\\
20.5 &  41.3\\
 \hline
 \end{tabular}
 \end{center}
 \label{tab3}
\end{table}

\begin{figure}
\centerline{\psfig{file=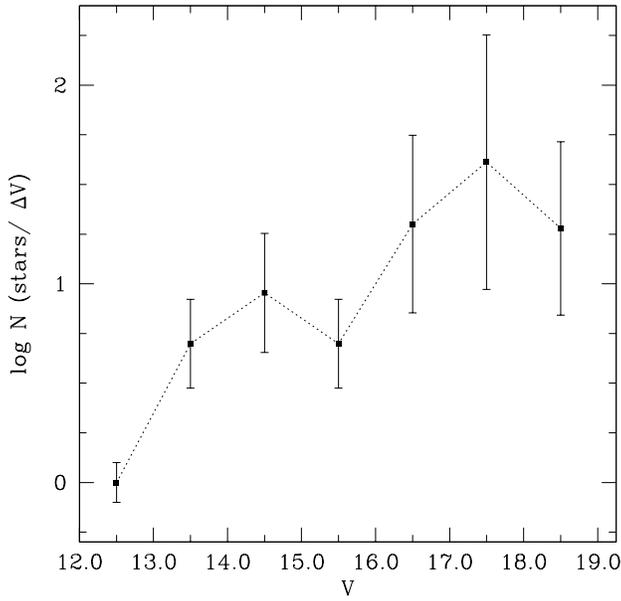,width=9cm}}
\caption{Cluster Luminosity Function.}
\label{fig6}
\end{figure}

Besides, we  assume that our comparison field provides a good estimate
of the contamination by field interlopers and is
valid across all the cluster surface.
The results are plotted in Fig.~\ref{fig6}.\\
Here, the completeness, background corrected stars counts
as a function of magnitude are shown together with their
Poisson uncertainties.
The LF is a rising function over the magnitude range
for which our completeness is larger than 50$\%$, except for
the last bin, at V larger than 18.
This seems to suggest that evaporation of low mass stars
has been important during NGC~1883 evolution, in agreement
with the findings in Sect.~8. Since this happens at the limit
of our photometry, a deeper study is mandatory to make stronger
our results.\\
\noindent
We note in addition the presence  of a statistically significant
depression at V $\sim$ 15.5, where
also the MS in the CMD (see Fig.~\ref{fig1}) does show a gap.
Such gaps are common in open clusters, and a variety
of explanations have been discussed over the years
for their occurrence (Rachford \& Canterna 2000).

\begin{figure}
\centerline{\psfig{file=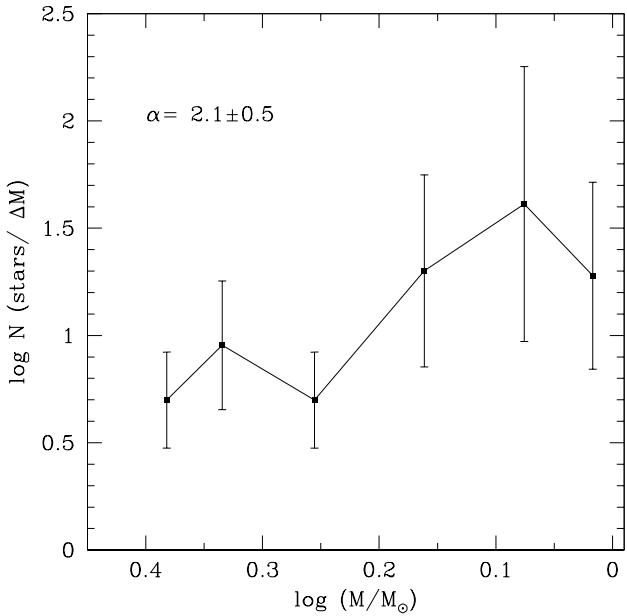,width=9cm}}
\caption{Cluster Mass Functions.}
\label{fig7}
\end{figure}
\noindent
Apparent magnitudes
were then transformed into the absolute ones
using the cluster apparent distance modulus (m-M)$_V$=14.3.
This allows us to compute the integrated absolute
magnitude of the cluster by adding up star fluxes.
We obtain M$_V$ = -2.4$\pm$0.2. This value has to be considered
as an upper limit, due to the possible presence of several
field stars which we are not able to completely remove.
\\

\noindent
Finally the $M_V$ distribution was converted into mass function (MF)
using Girardi et al. (2001) theoretical models.
The present day MF is shown in Fig.~\ref{fig7}. A linear fit through the points
yields 2.1$\pm$0.5 for the $\alpha$ index, close to
the Salpeter 2.35 value.
As for the LF, we notice a deficiency of less massive stars,
presumably due tidal mass loss.
The integration of the present day MF gives a current total mass
for NGC~1883 of about 240 M$_\odot$, 3 times smaller than
the mass at birth we derived in Sect.~7.
This lend further support to a scenario in which
substantial mass loss occurred as the cluster orbited
the Milky Way.

%
\section{Discussion and conclusions}
%

\begin{figure}
\centerline{\psfig{file=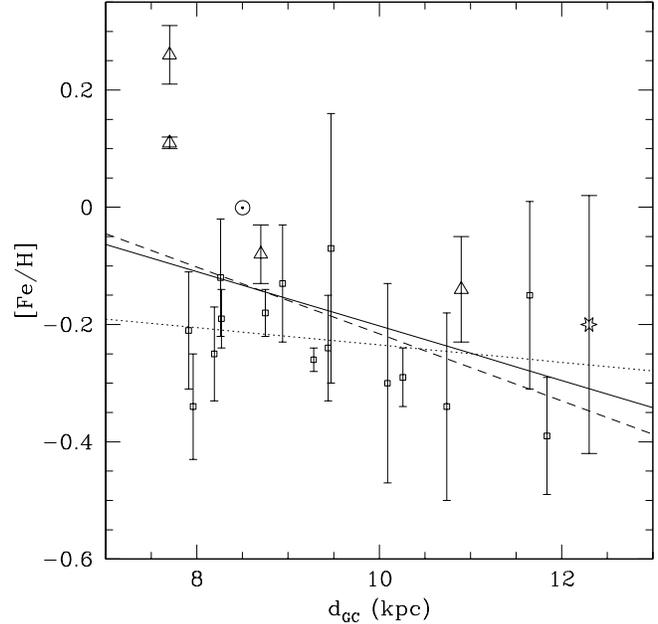,width=9cm}}
\caption{Galactic abundance gradient. Open squares
         are clusters from Friel et al. (2002),
         while open triangles are clusters
         found in the WEBDA database (see text for more details). 
         The open star is the value for NGC 1883.
         The position of the Sun (8.5,0) is indicated}
\label{fig8}
\end{figure}

In Figure~\ref{fig8} we plot the Galactic radial abundance
gradient as derived from Friel et al. (2002, hereafter F02)
from spectroscopic observations of a sample of open clusters (open squares).
We added also a sample of clusters not observed by F02 (open
triangles). We selected these objects from the 
WEBDA database\footnote{WEBDA database on open clusters
can be found at http://www.univie.ac.at/webda/}
with the condition of having metallicities spectroscopically determinated 
to be homogeneous with F02 and with this paper.
We plotted only clusters having age comparable with NGC 1883
(0.5$\leq$Age$\leq$2.0 Gyr, the same range used in F02
to identify the intermediate-age clusters sample). 
The whole sample defines an overall slope of -0.06$\pm$0.01 dex/kpc
(dashed line).

The open star represents NGC 1883, which follows the general trend very
nicely. In fact, the continuous line, which represents the radial
abundance gradient determined by including also NGC 1883
basically coincides with the previous one, having a slope of -0.05$\pm$0.01 dex/kpc.
But we must notice that the slope is mainly driven by the two 
inner most metal rich open clusters (the inner triangles). 
If we exclude these points, the
gradient is flat having a value of -0.01$\pm$0.01 dex/kpc
(the dotted line).
Considering also the dispersion of the data and the error
of the fit, we are led to the conclusion that Galactic gradient
is flat, at least outside the solar circle. Inside it
probably grows up to super-solar metallicity ([Fe/H]$\sim$+0.2). 
However this last point is not sure being based only on the
metallicities of two clusters, and it needs further investigations.

On the other hand, as discussed by F02, the oldest open clusters
(Age$>$4.0 Gr) formed in our Galaxy demonstrate a strong
gradient also outside the solar circle, 
as shown in Fig. 3 of F02. This is compatible with the
scenario of a primordial pollution of the interstellar matter
from which clusters were formed mainly in the inner 
regions of the Galaxy, following which the chemical enrichment
was extended to the outer regions.
The results was a present day homogenization
of the chemical content in the Galactic disk giving a flat gradient
as defined by the younger open clusters outside the solar circle.
However this scenario is not valid for the outer regions of the Milky
Way.\\
Carraro et al. (2004, hereafter C04), and Yong et al. (2005)
found the presence in the outer disk of
old open clusters (Age$>$4 Gyr) having a metallicity of about
-0.5 dex, well above the value of the gradient as defined by the
inner old clusters. This means that probably the outer disk
had a peculiar chemical evolution, probably driven by the contamination
of external objects (dwarf galaxies) cannibalized by the Milky.

On the other hand only few intermediate-age clusters were observed for metallicity
determination in the outer disk at medium-high resolution, 
the majority of them being located close to the Sun.

The addition of NGC~1883 indicates that the gradients keep flat 
for half a kpc more, suggesting a trend which further studies
can possibly confirm.

We finally note that the gradient exhibits quite a significant scatter. One
may wonder whether this solely depends on observational errors, or
whether this scatter reflects a true chemical inhomogeneity in the
Galactic disk.

Further observations are needed, especially in the
not well explored Galactic region between 14 and 22 kpc from the
Galactic center. We are actually engaged in such exploration, with
new results being presented in Carraro et al. (2007, submitted to AJ).

\section*{Aknowledgements}
This paper is based on data collected at the Mount Ekar Observatory,
Asiago (Italy). We wish to express our gratitude to the technical
staff of the Observatory.\\
In our investigation we made use of WEBDA.\\
Finally, the authors wish to thank the referee, Bruce Carney, for
useful comments.

%

\end{document}